\documentstyle[12pt,aps,prc,epsfig,preprint]{revtex}
\tightenlines
\begin{document}

\begin{titlepage}
\title{\vspace*{5mm}\bf
\Large The in-medium isovector  $\pi N$ amplitude from low energy
pion scattering}
\vspace{4pt}
\author{E. Friedman$^{a}$, M. Bauer$^{b}$, J. Breitschopf$^{b}$,
H. Clement$^{b}$, H. Denz$^{b}$, E. Doroshkevich$^{b}$,
A. Erhardt$^{b}$,  G.J. Hofman$^{c}$,
R. Meier$^{b}$, G.J.Wagner$^{b}$, G. Yaari$^{a}$\\
$^a${\it Racah Institute of Physics, The Hebrew University, Jerusalem 91904,
Israel\\}
$^b${\it Physikalisches Institut, Universit\"at T\"ubingen, 72076 T\"ubingen,
Germany\\}
$^c${\it TRIUMF, Vancouver, British Columbia, Canada V6T 2A3
and University of Regina, Regina, Saskatchewan, Canada S4S-0A2\\}}

\vspace{4pt}
\maketitle

\begin{abstract}
Differential cross sections for elastic scattering of 21.5 MeV 
positive and negative pions by Si, Ca, Ni and Zr have been measured as
part of a study of the pion-nucleus potential across threshold. The
`anomalous' repulsion in the $s$-wave term was observed, as is the
case with pionic atoms. The extra repulsion can be accounted for
by a chiral-motivated model where the pion decay constant
is modified in the medium. Unlike in pionic atoms, the anomaly 
cannot be removed
by merely  introducing an empirical on-shell energy dependence.
\newline
$PACS$: 12.39.Fe; 13.75.Gx; 21.30.Fe; 25.80.Dj
\newline
\newline 
Corresponding author: E. Friedman, 
Tel: +972 2 658 4667, \newline FAX: +972 2 658 6347, 
E mail: elifried@vms.huji.ac.il

\end{abstract}

\centerline{\today}
\end{titlepage}

A long-standing `anomalous' $s$-wave repulsion
 in the  pion-nucleus interaction at threshold, as
found in phenomenological analyses of strong interaction effects in
pionic atoms, has recently received  considerable attention
\cite{Wei00,KYa01,Fri02,FGa02,GGG02a,FGa03,KKW03,KKW03a,FGa04}
following the suggestion by Weise \cite{Wei00}
that it could  be expected, at least in the isovector channel, 
to result from a chirally motivated approach
where the pion decay constant 
becomes density dependent in the nuclear
medium. Very recently it was also argued \cite{KKW03,KKW03a} that the
energy dependence of the chirally expanded $\pi N$ isoscalar and
isovector amplitudes $b_{0}(E)$ and $b_{1}(E)$ respectively,
for zero-momentum {\it off shell} pions near threshold, could explain
the anomaly. In addition, the empirical {\it on shell} energy dependence 
 was shown
in Ref. \cite{FGa04} to be capable of explaining the anomaly
by imposing the minimal substitution requirement \cite{ETa82} of
$E \to E - V_{c}$, where $V_{c}$ is the Coulomb potential,
on the properly constructed pion optical potential
used in a large-scale fit to 100 pionic-atom data
across the periodic table.
The availability in recent years of strong interaction level shifts
and widths of `deeply bound' 1$s$ and 2$p$ pionic atom states
in isotopes of Pb and Sn
\cite{YHI96,GGK00,GGG02,SFG04}, obtained from the (d,$^3$He) reaction,
provides added impetus to the study of the $s$-wave pion-nucleus interaction
at threshold. 

In the present Letter we report on extending the study of the $s$-wave
term of the pion-nucleus potential by measuring the elastic scattering
of very low energy $\pi^+$ and $\pi^-$ on several nuclei. 
The purpose of this experiment is
to study the behaviour of the pion-nucleus potential across 
threshold into the scattering regime  and  to examine
if the above-mentioned anomaly is observed also above threshold. 
Of particular importance is
the question of whether the density dependence or the empirical energy
dependence, which remove the anomaly in pionic atoms, are required also
by the scattering data. In the scattering scenario, unlike in the atomic
case, one can study both charge states of the pion, thus increasing
sensitivities to isovector effects and to the energy dependence due
to the Coulomb interaction.

It is somewhat surprising to realize that at
kinetic energies well below 50 MeV there seems to be only one set of 
high quality
data available for both charge states of the pion obtained in
the same experiment,
namely, the data of Wright et al. \cite{Wri88} for 19.5 MeV pions on calcium.
Focusing  attention on the isovector channel, it is
desirable to include also nuclei with $N>Z$. 
Therefore the targets chosen for the
present experiment, which was carried out at 21.5 MeV pion kinetic energy, were 
 Si, Ca, Ni and Zr, where the last two
have an excess of neutrons. Natural isotopic mixtures have been used in all
cases and have been taken into account in the model calculations accordingly.

The experiment was performed at 21.5 MeV pion energy on the piE3 
channel at PSI \cite{piE3}, using the high resolution magnetic spectrometer LEPS
\cite{BKC90}. 
Particles were identified by time-of-flight relative
to the HF signal of the cyclotron and by time-of-flight within the 
spectrometer. Self
supporting targets were used in all cases. 
The beam was monitored by four decay-muon
telescopes  and by a downstream hodoscope \cite{BKC90}. The latter was also
used to determine the beam composition and profile.
The elastic scattering of muons
with the same momentum as the pions 
(except for the different energy losses in material such as beam line 
and spectrometer windows and plastic scintillators) 
was used to calibrate the absolute 
scale of cross sections and to check the overall validity of  data
reduction. This was achieved by comparing the measured 
angular distributions for
elastic scattering of muons with predictions made for 
Coulomb scattering from the known charge 
distributions of the target nuclei. Details on this method of muonic
normalization, which has frequently been used for low energy pion 
measurements with the LEPS spectrometer, can be found in 
Refs. \cite{BKC90,JMJ95}. 
Two measurements of elastic scattering
of muons were made for each measurement for pions: (i)Muons were recorded
in parallel with pions but at a slightly different location in the focal
plane, due to the different energy losses of muons and pions in the target
and in the various scintillators. (ii)Muons were recorded in designated
muon runs, where the spectrometer fields were adjusted to bring the muons
to the same location in the focal plane as the pions
in the proper pion runs. 
After correcting for effective target thickness and detector acceptance
we have obtained normalization constants common to all angles and to the two
types of measurements (i) and (ii) mentioned above, but slightly different
for $\mu^+$ and $\mu^-$, due to differences in the muon/pion ratios for
the different beams.
In this way we
could confirm the dependence of the acceptance of the spectrometer on
the position in the focal plane, which was determined separately by
scanning the magnetic fields.

\begin{figure}
\epsfig{file=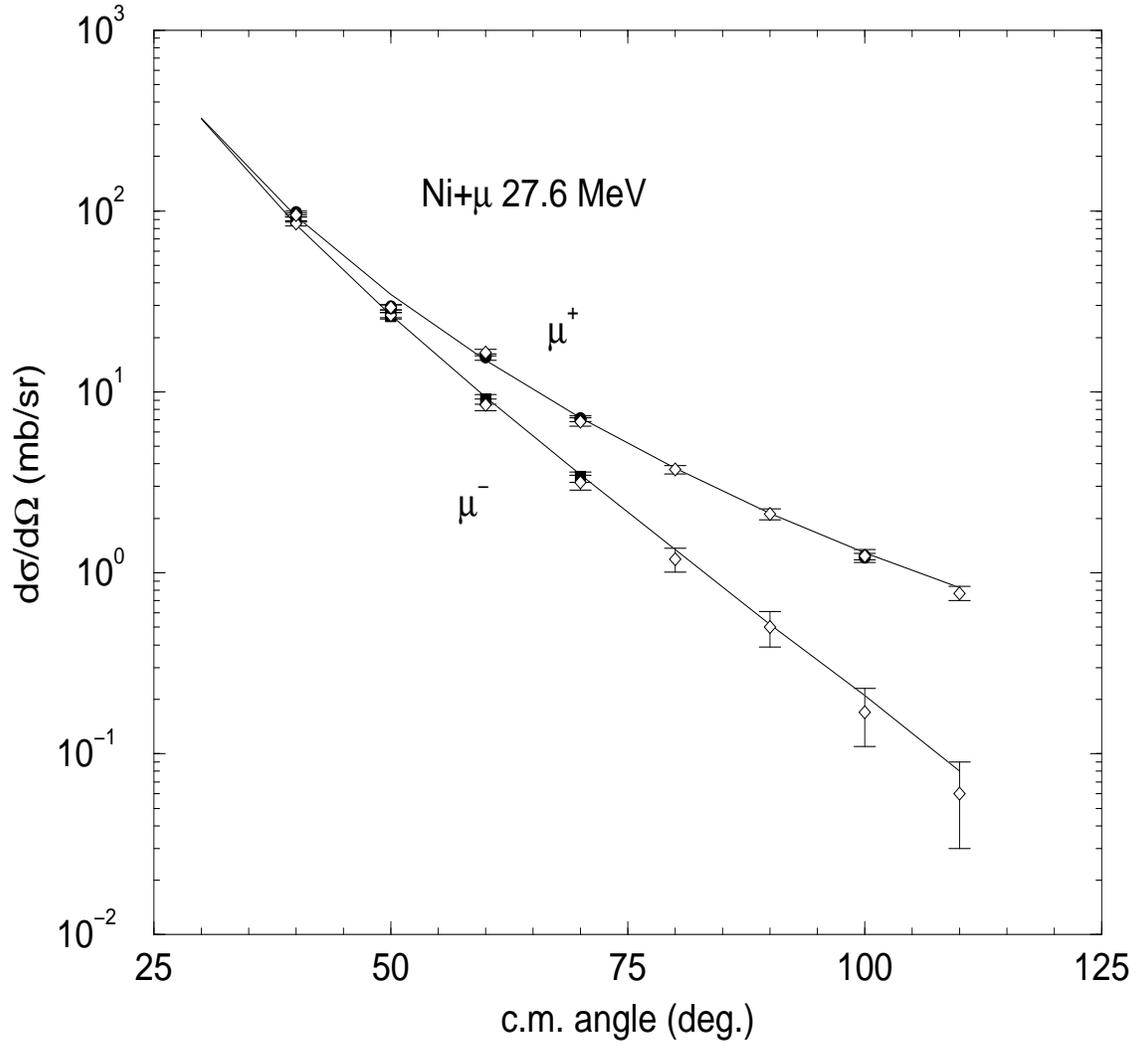, height=140mm,width=150mm}
\caption{Coulomb scattering of muons by Ni. Open symbols: from pion runs.
Filled symbols: from designated muon runs. 
Continuous curves are calculated Coulomb scattering for the finite size
charge distribution. Common normalization  constants have been
used, separately for all the $\mu^+$ points and all the $\mu^-$ 
points, see text.}
\label{fig:muons}
\end{figure}

Figure \ref{fig:muons} shows as an example comparisons between calculations
and measurements for the Coulomb scattering of muons by Ni. 
Open and filled symbols are for the two types of muon measurements 
(i) and (ii) mentioned above, respectively. The reliability of the
focal-plane position-dependence of the acceptance of LEPS
 is of major importance as the conclusions regarding the
pion-nucleus interaction (see below) rely exclusively on the {\it shape}
of the angular distributions. The points at 30$^\circ$ are not plotted
because different normalizations applied due to different settings
of the channel slits.

\begin{figure}
\epsfig{file=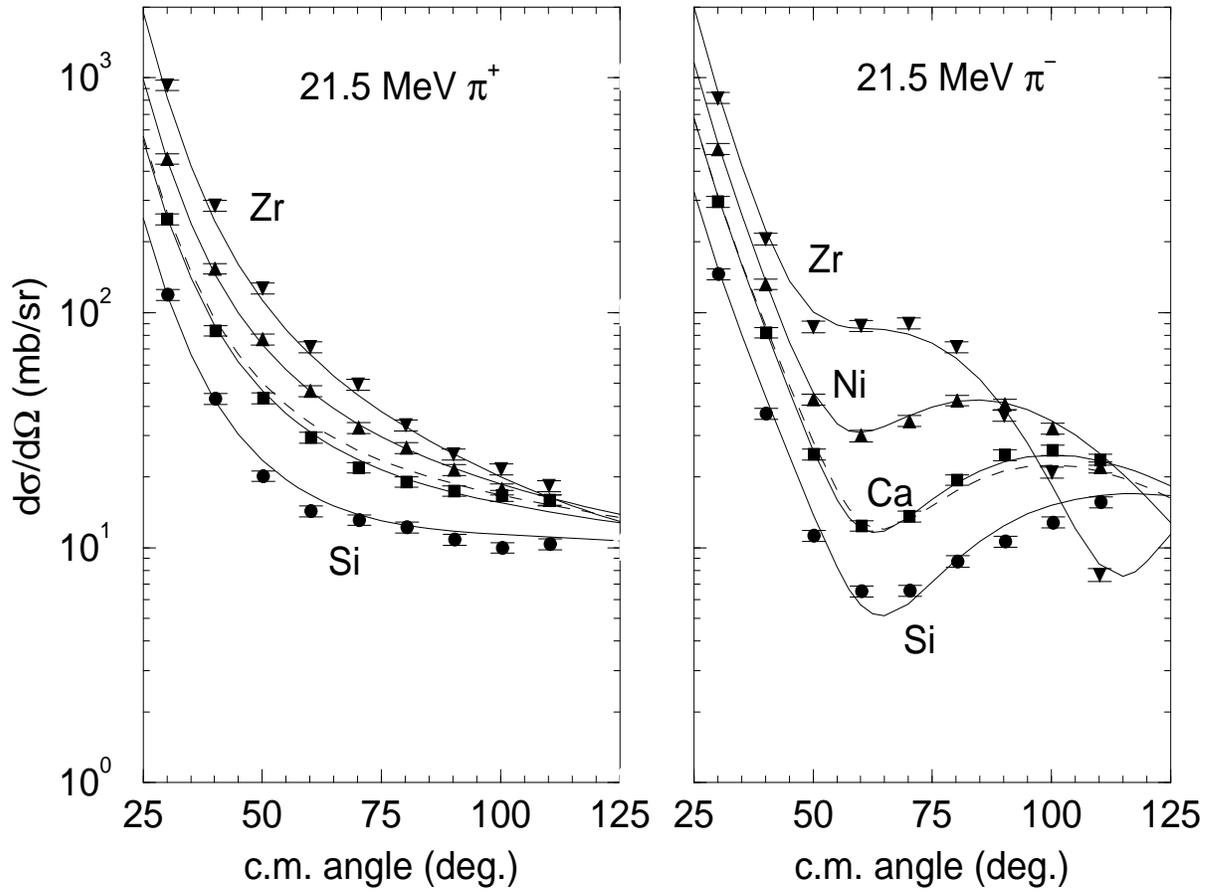, height=120mm,width=160mm}
\caption{Comparisons between experimental and calculated differential 
cross sections for elastic scattering of pions. Solid lines are
for the best-fit
optical potential (c') of Table \ref{tab:fits}, dashed lines are 
examples for potential (a'), see text.}
\label{fig:pions}
\end{figure}

Figure \ref{fig:pions} shows the experimental results for elastic
scattering of pions and predictions of  best-fit optical potentials.
Before discussing the implications of $\chi^2$ fits, 
 a comment is in order on values of errors.
(Full details of the experiment and the results will be published
elsewhere). The number of pion counts in the elastic scattering peaks
was usually greater than 3000 and the background in the spectrometer
was negligibly small, thus resulting in statistical errors of less than 2\%.
Most of the uncertainties in this experiment come from the monitoring of the
beam intensity and its composition and from the dependence of the acceptance
of the spectrometer on the position in the focal plane. The measurements
with muons provided stringent tests of the latter and careful analysis 
 did not reveal
any systematic effects. 
An estimated overall normalization error of 5\% was given in \cite{JMJ95}.
For the purpose of
$\chi^2$ fits we have added 5\% in quadrature to the statistical errors
for each point individually,
but in order to check the dependence of the derived parameter values on
the errors we have repeated the $\chi^2$ fits also for only 3\% added in
quadrature. Values of fit parameters and in particular values of the
isovector amplitude $b_1$ which is at the focus of the present work 
did not differ
between fits made with these two sets of errors.

The interaction of pions with target nuclei was described by the
Klein-Gordon equation with the standard potential 
due to  Ericson and Ericson \cite{EEr66} where double scattering,
absorption on two nucleons and angle-transformation terms have been
included \cite{FGa03,BFG97}. 
The potential is written as

\begin{equation} \label{equ:EE1}
2\mu V_{{\rm opt}}(r) = q(r) + \vec \nabla \cdot \alpha(r) \vec \nabla
\end{equation}
with its 
 $s$-wave term which is the prime concern
in the present work,  given by 


\begin{eqnarray} \label{equ:EE1s}
q(r) & = & -4\pi(1+\frac{\mu}{M})\{{\bar b_0}(r)[\rho_n(r)+\rho_p(r)]
  \pm b_1[\rho_n(r)-\rho_p(r)] \} \nonumber \\
 & &  -4\pi(1+\frac{\mu}{2M})4B_0\rho_n(r) \rho_p(r) 
\end{eqnarray}
where the $\pm$ sign refers to the pion $\mp$ charge states, respectively.
In these expressions $\rho_n$ and $\rho_p$ are the neutron and proton density
distributions normalized to the number of neutrons $N$ and number
of protons $Z$, respectively, 
$\mu$ is the pion-nucleus 
relativistic reduced mass and $M$ is the mass of the nucleon.
The function ${\bar b_0}(r)$ in Eq. (\ref{equ:EE1s})
is given in terms of the {\it local} Fermi
momentum $k_{\rm F}(r)$ corresponding to the isoscalar nucleon
density distribution:

\begin{equation} \label{equ:b0b}
{\bar b_0}(r) = b_0 - \frac{3}{2\pi}(b_0^2+2b_1^2)k_{\rm F}(r),
\end{equation}
where $b_0$ and $b_1$ are  the pion-nucleon isoscalar
and isovector effective scattering amplitudes, respectively.
The parameter $B_0$ represents $s$-wave absorption on pairs of nucleons.
The expressions for the $p$-wave term are given in Ref. \cite{FGa03,BFG97}
and will not be given explicitly here, except for its linear part, namely,

\begin{equation} \label{equ:alp}
\alpha(r) = 4\pi (1+\frac{\mu}{M})^{-1} \{c_0[\rho _n(r)
  +\rho _p(r)]  \pm c_1[\rho _n(r)-\rho _p(r)] \}  
{\rm ~~+~~quadratic~~terms}.
\end{equation}
Nuclear densities were obtained from charge densities and using for the
difference between neutron and proton rms radii either the 
results of relativistic mean field
calculations or values obtained from antiprotonic atoms, see 
Ref. \cite{FGa03} for details. Derived values of $b_1$ were insensitive
to assumptions on $\rho _n$ within those limits.

First attempts at parameter fits to the data using the above potential
ran into difficulties which could be traced to the two-nucleon absorption
in the $p$-wave term. Subsequent fits to the $\pi ^+$ data and the $\pi^ -$
data separately revealed the need to make the $p$-wave absorption 
parameter for $\pi^ -$
considerably larger than the corresponding parameter for  $\pi^ +$,
an effect not seen in earlier fits to only $\pi ^-$ data \cite{BBC94}.
This may be expected at 20 MeV since the effects of the (3,3) resonance
should depend on the energy, which is effectively higher for $\pi^ -$
than for $\pi^ +$. 
To avoid introducing isospin dependence into the quadratic $p$-wave term
we dropped it altogether
and made the parameters $c_0$ and $c_1$ complex.
Recall that only at threshold these cannot have imaginary terms.
The form of the $s$-wave part of the potential was kept initially 
as for pionic
atoms while its parameters were also varied in the fit process.

Least-squares fits
to the whole $\pi ^+$ and $\pi ^-$ data together produced reasonably 
good agreement between
calculation and experiment, with $b_0$ and the
real parts of  $c_0$ and $c_1$ close to the 
corresponding free $\pi N$ values
but with  $b_1$ 
significantly more repulsive than the
corresponding free $\pi N$ value of approximately $-$0.090 $m_\pi ^{-1}$.
The imaginary parts of $c_0$ and $c_1$ were 0.043$\pm$0.009 and
0.45$\pm$0.11 $m_\pi ^{-3}$, respectively. These
must be regarded as effective
because they now represent {\it all}  
non $s$-wave absorption processes. The $s$-wave part
of the potential which was kept as for pionic atoms showed a factor
of two reduction in the two-nucleon absorption 
Im$B_0$ compared to threshold.
This model is denoted below as (a).

With the value of $b_1$ found to be `anomalously' repulsive, as in
pionic atoms, we then applied the two mechanisms which have been found to
account for that anomaly in the pionic atoms case.
The first is that due to Weise \cite{Wei00}:
since $b_{1}$ in free-space is well approximated in lowest
chiral-expansion order by the Tomozawa-Weinberg expression
\cite{Tom66}

\begin{equation}
\label{equ:b1}
b_{1}=-\frac{\mu_{\pi N}}{8 \pi f^{2}_{\pi}}=-0.08~m^{-1}_{\pi} \,,
\end{equation}
then it can be argued that $b_{1}$ will be modified in the pion-nucleus
interaction if the pion decay constant
$f_\pi$ is modified in the medium. The square of this decay constant
is given, in leading order,
 as a linear function of the nuclear density,
$f_\pi ^{*2} = f_\pi ^2 - \rho \sigma / m_\pi ^2 $
with $\sigma$ the pion-nucleon sigma term.
This leads to a density-dependent isovector amplitude such that $b_1$ becomes
\cite{Fri02}
$b_1(\rho) = b_1(0) / (1-2.3\rho)$
for $\sigma $=50 MeV \cite{San02} and
with $\rho$ in units of fm$^{-3}$. This model is denoted by (b).
The second mechanism which has been successful in pionic atoms is to use 
the empirical on-shell energy dependence of the scattering amplitude $b_0$
and scattering volume $c_0$ within the minimal substitution requirement
$E \to E - V_{c}$ of \cite{FGa04,ETa82}, (model (c)). 
In additional fits  we have included both
mechanisms together, (model (d)).

Table \ref{tab:fits} (upper part) summarizes these fits, where
`mixed' refers to the model where the $s$-wave term of the potential is
of the conventional type 
whereas a linear form is used for the
$p$-wave part. In the lower part of the table results are given  for 
fits where the $s$-wave potential, too, was linear;
models (a') to (d') correspond to models (a) to (d) above, respectively.
By replacing the density-quadratic $s$-wave part 
by a complex linear part  we checked the dependence of the 
conclusions on the model used, 
particularly regarding the {\it in~medium} value of $b_1$,
which is found  to be essentially decoupled from the rest of
the potential.

The  obvious conclusions from the table are that (i)without the 
density dependence of $b_1$, its values differ from 
the free $\pi N$ value of $-$0.090 $m_\pi^{-1}$
by $\approx$ 3-4 standard deviations, (ii)with the density dependence
included,
the values of $b_1$ are in agreement with the free $\pi N$ value, and
(iii)the  inclusion of the empirical {\it on-shell} energy dependence
of $b_0$ and $c_0$ leads to  improved fits to the data.
The solid lines in
figure \ref{fig:pions} show the best  fit to the data obtained 
with the empirical energy dependence within the linear model (c'). 
Applying also the
density dependence of $b_1$ leads to very similar results.
The dashed lines show, as an example, results of model (a') for Ca
and it is evident that the agreement with the data is inferior to that
achieved with potential (c'), as is also evident from the table.
Note that at forward angles there is a  limited dependence of the
cross sections on the strong interaction potential and the agreement
between calculation and experiment essentially proves that the absolute
normalization is right. The information on the optical potential comes
from larger angles, thus it is the {\it shape} of  measured angular
distributions which is sensitive to the strong interaction.
If the data of Wright et al. \cite{Wri88} for 19.5 MeV pions scattered
from Ca are  also included in the analysis, then the resulting potential
parameters are unchanged but their $\pi ^-$ cross sections appear 
to be systematically
too small at larger angles by typically 10\%.

In conclusion, we have performed precision measurements of elastic scattering
of 21.5 MeV positive and negative pions on targets of Si, Ca, Ni and Zr 
with the aim of testing the `anomalous' $s$-wave repulsion observed
in pionic atoms. In particular, we focused on the question of whether
the mechanisms, which were shown recently to be capable of  removing 
the anomaly
in the atomic case, would be required also in the scattering regime
in order to reconcile the pion-nucleus interaction with the free
pion-nucleon interaction. It is found that (i) the in-medium
isovector amplitude $b_1$ is too repulsive by 3-4
standard deviations compared to the free $\pi N$ value
and (ii) that including the empirical {\it on-shell}
energy dependence of the scattering amplitude $b_0$ and the scattering volume
$c_0$  improves the fits to the data. 
However and  {\it unlike the case with
pionic atoms}, the energy dependence alone does {\it not} remove the extra
repulsion. Only with the inclusion of the chiral-motivated model where the pion decay constant
is modified in the medium, it is possible to reconcile the pion-nucleus interaction 
with the free pion-nucleon interaction, thus removing the anomaly in
the isovector channel.

\vspace{1cm}
We wish to acknowledge  many stimulating discussions with   A. Gal.
This work was supported in part by
the German minister of education and research (BMBF) under contracts 
06 TU 987I and  06 TU 201
and the Deutsche Forschungsgemeinschaft (DFG) through European Graduate
School 683 and Heisenberg Program.

\begin{table}
\caption{Results of $\chi ^2$ fits to the data using potentials
discussed in the text. The $b_1 (\rho )$ model is given by 
$b_1(\rho) = b_1(0) / (1-2.3\rho)$. When $b_1$ is complex (bottom half), the 
listed values refer to
its real part.}
\label{tab:fits}
\begin{tabular}{lcccc}
potential model & $b_0$ and $c_0$  &  $b_1$ model & $\chi ^2$ for 72 points &
$b_1$ ($m_\pi ^{-1}$) \\
\hline
mixed  (a) & fixed & fixed &273 & $-$0.125$\pm$0.015 \\
~~~~~~~~~(b)      &       & $b_1 (\rho)$& 273 & $-$0.098$\pm$0.008 \\
~~~~~~~~~(c)      & $b_0 (E)$ and $c_0 (E)$ & fixed & 197 & $-$0.117$\pm$0.011
 \\
~~~~~~~~~(d)      &                       & $b_1 (\rho)$& 215 & $-$0.087$\pm$0.010 \\
\hline
linear (a') & fixed & fixed &309 & $-$0.138$\pm$0.010 \\
~~~~~~~~(b')      &       & $b_1 (\rho)$& 259 & $-$0.096$\pm$0.006 \\
~~~~~~~~(c')      & $b_0 (E)$ and $c_0 (E)$ & fixed & 178 & $-$0.134$\pm$0.010
 \\
~~~~~~~~(d')      &                       & $b_1 (\rho)$& 214 & 
$-$0.093$\pm$0.007 \\
\end{tabular}
\end{table}

\end{document}